\documentclass[11pt, a4paper]{article}
\usepackage{latexsym,amsmath,amssymb,amsbsy,amsthm}

\newcommand{\beq}{\begin{equation}}
\newcommand{\feq}{\end{equation}}
\newcommand{\refn}[1]{(\ref{#1})}
\newcommand{\sol}[1]{SO(1,\,#1)}

\newcommand{\lvac}{\langle 0|}
\newcommand{\rvac}{|0\rangle}

\newcommand{\sst}{_{\star}}
\newcommand{\xco}[2]{(x_{#1},\,\ldots,\,x_{#2})}
\newcommand{\starp}[2]{\ph(x_{#1})\star\ldots\star\ph(x_{#2})}
\newcommand{\Tstar}[2]{T(\ph(x_{#1})\star\ldots\star\ph(x_{#2}))}
\newcommand{\Tgr}[2]{\lvac T(\ph(x_{#1})\star\ldots\star\ph(x_{#2}))\rvac}
\newcommand{\tph}{\tilde{\ph}}
\newcommand{\omk}{\omega(\vec k)}
\newcommand{\omq}{\omega(\vec q)}
\newcommand{\limred}[1]{\lim\limits_{p^0_{#1}\to\omega(\vec p_{#1})}}
\newcommand{\omp}[1]{\omega(\vec p_{#1})}

\newcommand{\thmn}{\theta^{\mu\nu}}
\newcommand{\km}{k_{\mu}}
\newcommand{\kn}{k_{\nu}}
\newcommand{\qm}{q_{\mu}}
\newcommand{\qn}{q_{\nu}}

\newcommand{\hx}{\hat x}

\newcommand{\hil}{\mathcal H}
\newcommand{\ph}{\varphi}

\newcommand{\der}{\partial}
\newcommand{\vx}{\vec x}
\newcommand{\vy}{\vec y}
\newcommand{\vk}{\vec k}
\newcommand{\vq}{\vec q}

\begin{document}

{\Large\bf LSZ reduction formula in many-dimensional \\
theory with space-space noncommutativity}

\medskip

\begin{quotation}\noindent
K.~V.~ Antipin,$^{1}$ M.~N.~Mnatsakanova,$^{2}$ and Yu.~S.~Vernov$^{3}$\vspace{-0.3\baselineskip}
\bigskip

\noindent$^1$\small{\em Department of Physics, Moscow State University, Moscow 119991, Russia. E-mail: kv.antipin@physics.msu.ru} \\
$^2$\small{\em Skobeltsyn Institute of Nuclear Physics, Moscow State University, Moscow 119992, Russia.} \\
$^3$\small{\em Institute for Nuclear Research, Russian Academy of Sciences, Moscow 117312, Russia.}

\bigskip\bigskip\bigskip\medskip

\noindent An analogue of the Lehmann---Symanzik---Zimmermann reduction formula is obtained for the case of noncommutative space-space theory. Some consequences of the reduction formula and Haag's theorem are discussed.
\bigskip
\medskip

\noindent Keywords: LSZ reduction formula;  noncommutative theory;  axiomatic quantum field theory.
\medskip

\noindent PACS numbers: 11.10.-z, 11.10.Cd

\end{quotation}
\vspace{-0.5\baselineskip}

\section{Introduction}
\mbox{}\vspace{-\baselineskip}

It is well known that in conventional quantum field theory the LSZ reduction formula allows effective calculation of the scattering amplitudes from the Green functions~\cite{LSZ}. In the present paper we analyze the applicability of this formula in the framework of noncommutative quantum field theory~(NCQFT). We will consider the case of a neutral scalar field in many-dimensional theory with space-space noncommutativity, so that the temporal variable commutes with the spatial ones. 

Let us consider the general case of $\sol{d}$-invariant theory with $d+1$ commutative coordinates~(including time) and an arbitrary even number $l$ of noncommutative ones. The commutation relations between $l$ noncommutative coordinates have the form
\beq\label{mnc}
\left[\hat x^i,\,\hat x^j\right]=i\theta^{ij},\quad i,\,j=1,\,\ldots,\,l,
\feq
where $\theta^{ij}$~--- real antisymmetric  $l\times l$ matrix. As we said, the rest $(d+1)$ variables commute with each other and all $\hx^j$ from~\refn{mnc}. 

In order to formulate the theory in commutative space-time, we use the Weyl ordered symbol~\cite{DN, Szabo} $\ph(x)$ of the noncommutative field operator $\Phi(\hat x)$:
\begin{equation}
\ph(x)=\frac1{(2\pi)^l}\int d^lk\int \textup{Tr}\,e^{ik(x-\hat x)}\Phi(\hat x),
\end{equation}
and the corresponding multiplication law $\ph_1\star\ph_2$ between the two symbols in the Weyl--Moyal--Groenewold form:
\begin{equation}\label{mp}
(\ph_1\star\ph_2)(x)=\left[e^{\frac{i}2\theta^{\mu\nu}\partial'_{\mu}\partial''_{\nu}}\ph_1(x')\ph_2(x'')\right]_{x'=x''=x}.
\end{equation}
Relation~\refn{mp} admits further generalization: for the symbols~(fields) taken at different points one can define twisted tensor product~\cite{Szabo, toward}
\begin{equation} \label{ttp}
\begin{split}
\ph (x_1) \star \ldots \star \ph (x_n) = \prod_{a<b} \,&\exp{\left
({\frac{i}{2} \, \theta^{\mu\nu} \, \frac{ {\partial}}{\partial
x^{\mu}_a} \, \frac{ {\partial}}{\partial x^{\nu}_b}} \right)} \,\ph
(x_1)  \ldots  \ph (x_n),\\
&a, b = 1,2, \ldots n.
\end{split}
\end{equation}
Thus the algebra of field operators is deformed, and it is not clear whether one can apply the standard LSZ formula for the noncommutative fields or not.
\newpage

\section{Commutation Relations for Creation and Annihilation Operators in NCQFT}\label{Weylrep}
\mbox{}\vspace{-\baselineskip}

As in conventional field theory, a free real scalar field in NCQFT admits a normal mode expansion:
\beq\label{decomp}
\begin{split}
\ph(x)=\ph^+&(x)+\ph^-(x),\\
\ph^{\pm}(x)=\frac1{(2\pi)^{(d+l)/2}}\int&\,\frac{d\vec k}{\sqrt{2\omk}}\left.\,e^{\pm ikx}\,a^{\pm}(\vec k)\right|_{k^0=\omk},
\end{split}
\feq
where $\omk=\sqrt{\vec{k}^2+m^2}$, $\vec{k}^2=\vec{k_c}^2+\vec{k}_{nc}^2$, $\vec{k_c}$--- commutative part of the $(d+l)$-dimensional vector $\vec{k}$, $\vec{k}_{nc}$--- noncommutative part of the same vector.

Let us obtain  commutation relations for the creation and annihilation operators $a^{\pm}$ directly from the assumption that the canonical quantization of a real scalar field in NCQFT is defined by  the relations
\beq\label{starcom}
\begin{split}
\left.[\ph(x)\star\der_0\ph(y)]\right|_{x^0=y^0}&=i\delta(\vx-\vy),\\
\left.[\ph(x)\star\ph(y)]\right|_{x^0=y^0}=0&,\quad\left.[\der_0\ph(x)\star\der_0\ph(y)]\right|_{x^0=y^0}=0.
\end{split}
\feq
Performing an inverse Fourier transform one can get the expression for $a^{\pm}$  from \refn{decomp}:
\beq
a^{\pm}(\vk)=\frac1{(2\pi)^{(d+l)/2}}\int\,d\vec x\,e^{\mp ikx}\,\left.\left[\sqrt{\frac{k_0}{2}}\ph(x)\mp\frac{i}{\sqrt{2k_0}}\der_0\ph(x)\right]\right|_{k_0=\omk}.
\feq
Let us take the operator product  $a^-(\vk)a^+(\vq)$ and  multiply it by $e^{\frac i2\thmn \km\qn}=e^{\frac i2\thmn i\km(-i)\qn}$. Expanding the phase factor in a series, we get
\beq
\begin{split}
e&^{\frac i2\thmn i\km(-i)\qn}\,a^-(\vk)a^+(\vq)=\sum_{n=0}^{\infty}\,\frac{(i/2)^n}{n!}\,\left(\thmn i\km (-i)\qn\right)^na^-(\vk)a^+(\vq)=\\
&=\frac1{(2\pi)^{(d+l)}}\,\int\int\,d\vx \,d\vy\,\sum_{n=0}^{\infty}\,\frac{(i/2)^n}{n!}\,\left(\thmn i\km (-i)\qn\right)^n\,e^{ikx}\,e^{-iqy}\times\\
&\quad\quad\quad\times\left(\sqrt{\frac{k_0}{2}}\ph(x)+\frac{i}{\sqrt{2k_0}}\der_0\ph(x)\right)\left(\sqrt{\frac{q_0}{2}}\ph(y)-\frac{i}{\sqrt{2q_0}}\der_0\ph(y)\right).
\end{split}
\feq
Next, let us replace momenta $k$ and $q$ with the derivatives, using the relation
\beq
(i\km(-i)\qn)^n\,e^{ikx}\,e^{-iqy}=\left(\der_{\mu}\der_{\nu}\right)^n\,e^{ikx}\,e^{-iqy},
\feq
and perform integration by parts, so that the derivatives act on the field $\ph$ in each term of the series. Thus we obtain $\star$-product of the field operators:
\beq\label{et1}
\begin{split}
&e^{\frac i2\thmn \km\qn}\,a^-(\vk)a^+(\vq)=\frac1{(2\pi)^{(d+l)}}\,\int\int\,d\vx \,d\vy\,e^{ikx}\,e^{-iqy}\times\\
&\times\left(\sqrt{\frac{k_0}{2}}\ph(x)+\frac{i}{\sqrt{2k_0}}\der_0\ph(x)\right)\star\left.\left(\sqrt{\frac{q_0}{2}}\ph(y)-\frac{i}{\sqrt{2q_0}}\der_0\ph(y)\right)\right|_{k_0=\omk,\,q_0=\omq}.
\end{split}
\feq
Similarly,
\beq\label{et2}
\begin{split}
&e^{\frac i2\thmn \qm\kn}\,a^+(\vq)a^-(\vk)=\frac1{(2\pi)^{(d+l)}}\,\int\int\,d\vx \,d\vy\,e^{ikx}\,e^{-iqy}\times\\
&\times\left(\sqrt{\frac{q_0}{2}}\ph(y)-\frac{i}{\sqrt{2q_0}}\der_0\ph(y)\right)\star\left.\left(\sqrt{\frac{k_0}{2}}\ph(x)+\frac{i}{\sqrt{2k_0}}\der_0\ph(x)\right)\right|_{k_0=\omk,\,q_0=\omq}.
\end{split}
\feq
Now, taking the fields $\ph(x)$ and $\ph(y)$ at equal moments of time $x^0=y^0$ and subtracting~\refn{et2} from~\refn{et1}, with the use of~\refn{starcom} we obtain:
\beq
e^{\frac i2\thmn \km\qn}\,a^-(\vk)a^+(\vq)-e^{\frac i2\thmn \qm\kn}\,a^+(\vq)a^-(\vk)=\delta(\vk-\vq),
\feq
or, in a more convenient form:
\beq\label{acom1}
a^-(\vk)a^+(\vq)=e^{-i\thmn \km\qn}\,a^+(\vq)a^-(\vk)+e^{-\frac i2\thmn \km\qn}\,\delta(\vk-\vq).
\feq
In the same way we get
\beq\label{acom2}
a^{\pm}(\vk)a^{\pm}(\vq)=e^{i\thmn \km\qn}\,a^{\pm}(\vq)a^{\pm}(\vk).
\feq
Commutation relations~\refn{acom1} and~\refn{acom2} are equivalent to the ones obtained in~\cite{twPoin} from general group-theoretical considerations involving the twisted Poincar\'e symmetry.

\section{Analogue of the LSZ reduction formula for space-space NCQFT}
\mbox{}\vspace{-\baselineskip}

In~\cite{toward} it was proposed that the expression for the noncommutative Wightman functions has the following form:
\beq\label{ncW}
W\sst(x_1,\ldots,\,x_n)=\lvac\starp{1}{n}\rvac,
\feq
where $\star$-product of fields taken at independent points is given by~\refn{ttp}.

In accordance with~\refn{ncW} we suppose that the noncommutative Green functions are
\beq\label{GrF}
G\sst\xco{1}{n}=\Tgr{1}{n},
\feq
where we defined time-ordered $\star$-product of fields as straightforward generalization of the usual $T$-product:
\beq
\begin{split}
T(\ph_1(x_1)\star\ldots\star\ph_n(x_n))&=\ph_{\sigma_1}(x_{\sigma_1})\star\ldots\star\ph_{\sigma_n}(x_{\sigma_n}),\\
x_{\sigma_1}^0>x_{\sigma_2}^0&>\ldots>x_{\sigma_n}^0.
\end{split}
\feq

Below we extend the classical proof of the LSZ formula~\cite{LSZ, BD, IZ} to the case of space-space NCQFT.

Let us single out the variable $x_1$ and consider the expression
\beq
\limred{1}\,(p_1^2-m^2)\int dx_1^0\,d\vec x_1\,e^{-ip_1x_1}\Tgr{1}{n}.
\feq
Dividing integration over $dx_1^0$ into three parts
\beq\label{tpart}
\int\,(\ldots)\,dx_1^0=\int_{-\infty}^{-\tau}\,(\ldots)\,dx_1^0+\int_{-\tau}^{\tau}\,(\ldots)\,dx_1^0+\int_{\tau}^{+\infty}\,(\ldots)\,dx_1^0,
\feq
we denote the summands as $I_1(\tau)$, $I_2(\tau)$, and $I_3(\tau)$ respectively.

Using  expression $(p_1^2-m^2)e^{-ip_1x_1}=(\square_1-m^2)e^{-ip_1x_1}$ and performing integration by parts, we get:
\beq\label{redmain}
\begin{split}
&I_1(\tau)=\int\,d\vec x_1\,e^{i\vec p_1\vec x_1+i\omega(\vec p_1)\tau}\lvac\Tstar{2}{n}\star\\
\star(i&\omp{1}-\frac{\der}{\der\tau})\ph(-\tau,\,\vec x_1)\rvac-\int_{-\infty}^{-\tau}\,dx_1^0\,\int\,d\vec x_1\,e^{i\vec p_1\vec x_1-i\omega(\vec p_1)x_1^0}\times\\
&\quad\quad\quad\times\lvac\Tstar{2}{n}\star(\square_1-m^2)\ph(x_1)\rvac.
\end{split}
\feq
Here $\square_1\equiv\frac{\der^2}{\der (x_1^1)^2}+\frac{\der^2}{\der (x_1^2)^2}+\frac{\der^2}{\der (x_1^3)^2}-\frac{\der^2}{\der (x_1^0)^2}$, and $\tau$ is taken sufficiently large so that the  permutation of $\ph(x_1)$ to the last position on the right is possible.

Next, we  use the Fourier-expression for $\ph(-\tau,\,\vec x_1)$:
\beq\label{specred}
\ph(-\tau,\,\vec x_1)=\frac1{(2\pi)^{(d+l)/2}}\int\,dk_0\,d\vec k\,e^{-ik_0\tau}\,e^{-i\vec k\vec x_1}\tph(k).
\feq
All derivatives in the $\star$-product will act on the factor $e^{-i\vec k\vec x_1}$ in the Fourier expansion of $\ph(x_1)$. Therefore, additional factor $N(k_{nc})$ will appear. Note that $N(k_{nc})$ depends only on the noncommutative part of $\vk$.

Let us also take into account the asymptotic representation for the field $\ph$:
\beq
\lim\limits_{t\to -\infty}\int\,dk_0\,e^{it(k_0-\omk)}\tph(k)=\frac1{\sqrt{2\omk}}a^+_{in}(\vec k).
\feq
Taking the limit $\tau\to\infty$, we obtain:
\beq
\begin{split}
&I_1=\lim\limits_{\tau\to\infty}I_1(\tau)=i(2\pi)^{(d+l)/2}\int\,dk_0\,(k_0+\omp{1})\delta(k_0-\omp{1})\times\\
\times&\lvac\Tstar{2}{n}N(p_{1,\,nc})\frac{a^+_{in}(\vec p_1)}{\sqrt{2\omp{1}}}\rvac=\\
=i&(2\pi)^{(d+l)/2}\sqrt{2\omp{1}}N(p_{1,\,nc})\lvac\Tstar{2}{n}a^+_{in}(\vec p_1)\rvac.
\end{split}
\feq
In this limit the second term of the expression~\refn{redmain} is equal to null.

Similar calculations for $I_3(\tau)$ will give:
\beq\label{i3red}
I_3=i(2\pi)^{(d+l)/2}\sqrt{2\omp{1}}N(p_{1,\,nc})\lvac a^+_{out}(\vec p_1)\Tstar{2}{n}\rvac=0.
\feq
As to the second summand in~\refn{tpart}, $I_2(\tau)$ can be presented as
\beq
\begin{split}
\int_{-\infty}^{\infty}\,&dx_1^0\,e^{-ip_1^0x_1^0}\,\chi(x_1^0,\,\tau)\,F(x_1),\\
&\chi(x_1^0,\,\tau)=\begin{cases}
1,\quad |x_1^0|\leqslant\tau;\\
0,\quad |x_1^0|>\tau.
\end{cases}
\end{split}
\feq
The integrand contains a generalized function with  compact support, so its Fourier-transform is a smooth function and doesn't have a pole. For this reason
\beq
\limred{1}\,(p_1^2-m^2)\int_{-\tau}^{\tau} dx_1^0\,\int\,d\vec x_1\,e^{-ip_1x_1}\Tgr{1}{n}=0.
\feq

Following similar limiting procedure  over $x_1$ and $x_2$ consecutively, we obtain
\beq
\begin{split}
&\limred{2}\limred{1}(p_2^2-m^2)(p_1^2-m^2)\int\,dx_1\,\int\,dx_2\,e^{-ip_2x_2-ip_1x_1}\times\\
\times&\Tgr{1}{n}=\left[i(2\pi)^{(d+l)/2}\right]^2\sqrt{2\omp{1}}\sqrt{2\omp{2}}\times\\
\times & N(p_{2,\,nc})N(p_{1,\,nc})\lvac\Tstar{3}{n}a^+_{in}(\vec p_2)a^+_{in}(\vec p_1)\rvac.
\end{split}
\feq
Now let us replace the secong procedure~(over $x_2$) with the one corresponding to transition to the bottom sheet of the mass hyperboloid, that is $\lim\limits_{p_2^0\to -\omp{2}}(p_2^2-m^2)$. Making use of the asymptotic representation
\beq
\lim\limits_{t\to \pm\infty}\int\,dk_0\,e^{it(k_0+\omk)}\tph(k)=\frac1{\sqrt{2\omk}}a^-_{in(out)}(-\vec k),
\feq
we obtain:
\beq\label{finlim}
\begin{split}
&\lim\limits_{p_2^0\to -\omp{2}}\limred{1}(p_2^2-m^2)(p_1^2-m^2)\int\,dx_1\,\int\,dx_2\,e^{-ip_2x_2-ip_1x_1}\times\\
\times&\Tgr{1}{n}=
\left(i(2\pi)^{(d+l)/2}\right)^2\sqrt{2\omp{1}}\sqrt{2\omp{2}}\times\\
\times &[N(p_{2,\,nc})N(p_{1,\,nc})\lvac a_{out}^-(-\vec p_2)\Tstar{3}{n}a_{in}^+(\vec p_1)\rvac-\\
-&\tilde N(p_{2,\,nc})\tilde N(p_{1,\,nc})\lvac \Tstar{3}{n}a^-_{in}(-\vec p_2)a^+_{in}(\vec p_1)\rvac].
\end{split}
\feq
The first term here is the contribution of $I_3(\tau)$, which is not equal to zero in this limit. Let us make the substitution $\vec p_2\to-\vec p_2$. We consider the scattering processes in which $\vec p_1$ --- incoming momentum, $\vec p_2$ --- outcoming momentum, and $\vec p_1\ne\vec p_2$. In accordance with~\refn{acom1} we can commute $a^-_{in}(\vec p_2)$ and $a^+_{in}(\vec p_1)$ in the second term of~\refn{finlim} so that $a^-_{in}(\vec p_2)$ can act on the vacuum state and give null.

We can repeat the above-mentioned procedure $n$ times --- until nothing is left under the time-ordered $\star$-product. Now the additional factor $N(p_{1,\,nc})\times\ldots\times N(p_{n,\,nc})$ can be expressed in explicit form: each derivative $\der_{\mu}$ in~\refn{ttp} should be replaced with $ip_{\mu}$, and we have:
\beq\label{dist}
\begin{split}
N(p_{1,\,nc})\times\ldots\times N(p_{n,\,nc})=&\exp\left.\left[-\frac i2\, \theta_{\mu\nu}\sum_{a<b}\,p^{\mu}_{a}p^{\nu}_{b}\right]\right|_{-p_{out}},\\
& a,\,b=1,\,\ldots,\,n,
\end{split}
\feq
where $\left.\right|_{-p_{out}}$ means that outcoming momenta should be taken with the minus sign~(as the result of the substitution $\vec p\to-\vec p$ we made earlier).

The final expression for the scattering amplitude:
\beq\label{LSZnc}
\begin{split}
&\lvac a_{out}^-(\vec p_1)\,\ldots\,a_{out}^-(\vec p_k)\,a_{in}^+(\vec p_{k+1})\,\ldots\,a_{in}^+(\vec p_{n})\rvac=\left[\frac1{i(2\pi)^{(d+l)/2}}\right]^n\times\\
\times &\,\exp\left.\left[\frac i2\, \theta_{\mu\nu}\sum_{a<b}\,p^{\mu}_{a}p^{\nu}_{b}\right]\right|_{-p_{out}}\,\prod_{j=1}^{n}\,\frac{p_j^2-m^2}{\sqrt{2\omp{j}}}\,G\sst(-p_1,\,\ldots,\,-p_k,\,p_n,\,\ldots,\,p_{k+1}),
\end{split}
\feq
where $G\sst(p_1,\,\ldots,\,p_n)$ --- Fourier transform of the noncommutative Green function:
\beq\label{Fourie}
G\sst(p_1,\,\ldots,\,p_n)=\int\,dx_1\ldots dx_n\,\exp\left[-i\sum_{j=1}^{n}\,p_jx_j\right]\,G\sst(x_1,\,\ldots,\,x_n).
\feq

Relation~\refn{LSZnc} is a noncommutative analogue of the LSZ reduction formula. This result corresponds to the one obtained in~\cite{LSZmx}, authors of which didn't use the $\star$-product between the fields taken at different points and considered the Green functions with the usual time-ordered product of noncommutative fields. The difference between the two results is the additional phase-factor~\refn{dist} due to the chosen form of the Green function~\refn{GrF}.

\section{Consequences}
\mbox{}\vspace{-\baselineskip}

Now we can extend to NCQFT the considerations that were originally proposed in~\cite{classth} for the case of commutative theory.

Suppose that we have two noncommutative $\sol{d}$-invariant theories on Hilbert spaces $\hil_1$ and $\hil_2$ respectively, related by a unitary transformation. Let $\ph_1$ and $\ph_2$ be two irreducible sets of field operators defined in $\hil_1$ and $\hil_2$.  Let $< p'_1,\ldots,\,p'_n | p_1,\dots,\, p_m >_{i},\,i=1,2$ be inelastic scattering amplitudes of the process $m \to n$ for the fields $\ph_1$ and $\ph_2$ respectively. In accordance with the reduction formula~\refn{LSZnc}
\begin{equation} \label{red}
\begin{split}
&< p'_1,\ldots,\,p'_n | p_1,\dots,\, p_m >_{i} \,\sim\\
\sim\int \,d\,x_1 \ldots d\,x_{n+m} \, &\exp\{i \, (-p_1 \,x_1 -\ldots- p_m \,x_m+p'_1 \,x_{m+1} + \ldots + p'_n \,x_{n+m})\}\times\\
\times &\prod_{j =1} ^{n+m} \, (\square_{j} - m ^{2}) \, \lvac T(\varphi_{i} \,(x_1)\star \, \ldots \,\star \varphi_{i} \, (x_{n+m})) \rvac,\\
&\quad\quad\quad\quad\quad\quad i=1,\,2.
\end{split}
\end{equation}

Let us also take into account the results obtained for the generalized Haag's theorem in the context of noncommutative theory~\cite{vest,nucl}. Namely, it was shown that in two $\sol{d}$-invariant theories, related by a unitary transformation, the two-, three, \ldots,  $d+1$-point Wightman functions coincide:
\beq\label{wicoi}
\begin{split}
\lvac \varphi_{1} \,(x_1)\star \, \ldots \,\star \varphi_{1} \, (x_{s}) \rvac&=\lvac \varphi_{2} \,(x_1)\star \, \ldots \,\star \varphi_{2} \, (x_{s}) \rvac,\\
2\leqslant& s\leqslant d+1.
\end{split}
\feq
From~\refn{red} and~\refn{wicoi} it follows that the amplitudes $< p'_1,\ldots,\,p'_n | p_1,\dots,\, p_m >_1$ and $< p'_1,\ldots,\,p'_n | p_1,\dots,\, p_m >_2$ coincide in the two theories if 
\beq
m+n\leqslant d+1.
\feq


\begin{thebibliography}{0}    

\bibitem{LSZ} H. Lehmann., K. Symanzik, W. Zimmermann, {\it Nuovo Cimento} {\bf 1}, 205 (1955).

\bibitem{DN} M. R. Douglas, N. A. Nekrasov, {\it Rev. Mod. Phys.} {\bf 73}, 977 (2001), [arXiv:hep-th/0106048].

\bibitem{Szabo} R. J. Szabo, {\it Phys. Rep.} {\bf 378}, 207 (2003), [arXiv:hep-th/0109162].

\bibitem{toward} M. Chaichian, M. N. Mnatsakanova, K. Nishijima, A. Tureanu, and Yu. S. Vernov, {\it J. Math. Phys.} {\bf 52}, 032303 (2011), [arXiv:hep-th/0402212].

\bibitem{twPoin} A.P. Balachandran, T.R. Govindarajan, G. Mangano, A. Pinzul, B.A. Qureshi, and S.
Vaidya, {\it Phys. Rev. D} {\bf 75}, 045009 (2007), [arXiv: hep-th/0608179].

\bibitem{BD} J.~D.~Bjorken and S.~D.~Drell, {\it Relativistic Quantum Mechanics Vol. 2} (McGraw-Hill Inc.,  1964).

\bibitem{IZ} C.~Itzykson, J.-B.~Zuber,
{\it Quantum Field Theory Vol. 1} (McGraw-Hill Inc., 1980).

\bibitem{LSZmx} A.P. Balachandran, P. Padmanabhan, A. R. de Queiroz {\it Phys. Rev. D} {\bf 84}, 065020 (2011), [arXiv: hep-th/1104.1629].

\bibitem{classth} M. Chaichian, M. Mnatsakanova, A. Tureanu, Yu. Vernov, {\it Classical Theorems in Noncommutative Quantum Field Theory},  preprint [arXiv: hep-th/0612112].

\bibitem{vest} K. V. Antipin, M. N. Mnatsakanova, Yu. S. Vernov,  {\it Moscow Univ. Phys. Bull.} {\bf 66}, 349 (2011), [arXiv: hep-th/1102.1195].

\bibitem{nucl} K. V. Antipin, M. N. Mnatsakanova, Yu. S. Vernov, {\it Phys. of Atom. Nucl. } {\bf 76}, 965 (2013), [arXiv: hep-th/1202.0995 ].


\end{thebibliography}
\end{document}